\documentclass[aps,prl,reprint,superscriptaddress,amsmath,amssymb,longbibliography,floatfix]{revtex4-2}

\usepackage[utf8]{inputenc}
\usepackage[T1]{fontenc}
\usepackage{lmodern}
\usepackage{graphicx}
\usepackage{bm}
\usepackage{xcolor}
\usepackage{hyperref}
\usepackage{dsfont}


\newcommand{\ii}{\mathrm{i}}

\newcommand{\Id}{\mathds{1}}
\newcommand{\Lmat}{\mathcal{L}}   
\newcommand{\Umat}{\mathcal{U}}   
\newcommand{\Mmat}{\mathcal{M}}   

\begin{document}

\title{Floquet Engineering of Momentum Gaps in Magnonic Time Crystals: Slow-Drive Amplification and In-Gap Propagation}

\author{Pawe\l{} Gruszecki}
\email{gruszecki@amu.edu.pl}
\affiliation{Institute of Spintronics and Quantum Information, Faculty of Physics and Astronomy, Adam Mickiewicz University, Uniwersytetu Pozna\'nskiego 2, 61-614 Pozna\'n, Poland}

\author{Victor Pacheco-Pe\~na}
\affiliation{School of Mathematics, Statistics and Physics, Newcastle University, Newcastle upon Tyne NE1 7RU, United Kingdom}

\date{\today}

\begin{abstract}
Photonic time crystals host momentum gaps in which waves grow exponentially but cannot propagate. We establish their magnonic counterpart in a thin film with periodically switched anisotropy, deriving its Floquet band structure and confirming it micromagnetically. Gaps require a precession-ellipticity contrast. The waveform controls high-order gaps: a single 1~GHz switching drive amplifies a comb of spin-wave bands, from 2.5 to 4.5~GHz, above Gilbert losses even with smoothed edges, unlike sinusoidal modulation. With the Dzyaloshinskii--Moriya interaction, amplified packets propagate inside the gap while growing.
\end{abstract}

\maketitle

A spatially uniform temporal modulation of a medium conserves the wave vector but changes the frequency, giving rise to temporal refraction and reflection~\cite{Morgenthaler1958,Galiffi2023,Moussa2023, PachecoPena2025Temporal, Galiffi2025Electrodynamics}.
Temporally periodic modulation opens momentum gaps (MGs), in which Floquet modes grow or decay exponentially~\cite{Cassedy1963_PartI,Cassedy1967_PartII,ZuritaSanchez2009,Lustig2018,Asgari2024, Meissner1918,KoutserimpasFleury2018}; 
MGs have been demonstrated experimentally, for instance, in microwave transmission lines~\cite{ReyesAyona2015,Xiong2025,Zhu2025}, while recent theoretical studies suggest routes toward their realization at optical frequencies~\cite{Wang2025_bandgaps,Garg2026}.


In magnonics, temporal modulation has enabled spin-wave frequency conversion~\cite{Rezende1967,Auld1967}, time reversal~\cite{Chumak2010}, and space--time-resolved temporal refraction~\cite{Schultheiss2021}, and the precession ellipticity has been identified as the magnonic temporal impedance, which governs time reflection into antimagnons and spin-wave amplification at temporal interfaces~\cite{SobuckiGruszecki2026}.
Floquet magnons have been studied mainly in the nonlinear regime, where the periodicity arises from the magnetization dynamics itself, e.g., in vortices and synthetic antiferromagnets~\cite{Heins2026,Devolder2026_TimeResolved,DevolderKim2026,Philippe2025}, whereas the band structure of a driven space--time crystal was not resolved~\cite{Trager2021}.

By contrast, under externally imposed weak harmonic modulation, an MG corresponds to parametric amplification~\cite{Khurgin2024_Parametric,Asgari2024}.
In magnonics, this mechanism is driven by parametric pumping, studied since the late 1950s~\cite{Schlomann1960,Lvov1994,Melkov1996}.
A pump at angular frequency $\omega_p$ couples spin-wave modes with opposite wave vectors $\pm k$ near the resonance condition $\omega_k=\omega_p/2$, where $\omega_k$ is the angular frequency of the unmodulated spin-wave mode with wave vector $k$.
For small spin-wave amplitudes, the dynamics remains linear with time-periodic coefficients.
Parametric pumping enables phase conjugation of spin-wave packets~\cite{Melkov2000,Melkov2001,Serga2005}, as well as spin-wave excitation and amplification down to the nanoscale~\cite{Bracher2014,Chen2017,Verba2017,Verba2018,Deka2022,AuFripp2023,Rivard2025,Nikolaev2025}.
Most of these studies address the fundamental parametric resonance, where the spin-wave frequency is half the modulation frequency.
Higher-order parametric resonances have also been reported in nonlinear ferromagnetic resonance (FMR), where large-amplitude uniform precession itself provides the modulation~\cite{Bauer2015_LowField}, and predicted for a single resonant mode~\cite{Grimaldi2026}.
These studies, however, did not address propagating spin-wave signals.
Even when Floquet theory is applied, the analysis focuses on parametric instabilities of individual modes~\cite{Bauer2015_LowField}.
This leaves open how MGs emerge in the Floquet band structure of propagating spin waves under an externally imposed, spatially uniform periodic modulation.
A related question is how the modulation waveform---from abrupt switching through smoothed transitions to a sinusoid---controls higher-order gaps.
Can these gaps enable a low-frequency drive to amplify propagating spin waves at frequencies well above half the drive frequency, with gain sufficient to overcome Gilbert damping?
These questions motivate our study.

Here we establish a magnonic time crystal in a film with periodically switched perpendicular magnetic anisotropy (PMA), deriving its Floquet bands and verifying them with micromagnetic simulations. 
As it will be shown, ellipticity contrast opens MGs, while the waveform and accumulated phase control higher-order gaps, enabling a slow drive to amplify several spin-wave bands above Gilbert losses. 
With interfacial Dzyaloshinskii--Moriya interaction (DMI), amplified wave packets drift rigidly inside the MGs of a single band at a DMI-set velocity, whereas in photonic time crystals single-band gap modes do not propagate~\cite{Lee2026_EnergyVelocity}, and in-gap transport has relied on coupling between different bands~\cite{MultiMode2026}.

\begin{figure}[!tb]
    \centering
    \includegraphics[ width=\columnwidth,height=0.46\textheight,keepaspectratio]{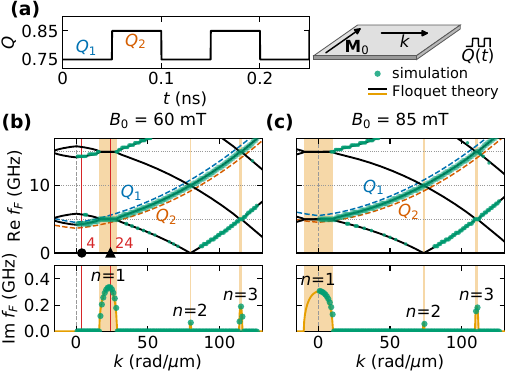}
    \vspace{-20pt}
    \caption{
(a) Reduced anisotropy $Q(t)$ switched between $Q_1=0.75$ and $Q_2=0.85$ with period $T=0.1$~ns ($f_{\mathrm{mod}}=10$~GHz, 50\% duty cycle); right: geometry, $\mathbf{M}_0\parallel\mathbf{B}_0\parallel\hat{y}$, $\mathbf{k}\parallel\hat{x}$.
(b),(c) Real (top) and imaginary (bottom) parts of the Floquet quasifrequency $f_F$ versus $k$ for $B_0=60$~mT (b) and $85$~mT (c): Floquet theory (black: $\operatorname{Re}f_F$; gold: growing branch of $\operatorname{Im}f_F$) and micromagnetic simulations, one per $k$ (green symbols, $\alpha=10^{-4}$).
Dashed: unmodulated dispersions for $Q_1$ (blue) and $Q_2$ (orange); dotted: $nf_{\mathrm{mod}}/2$; shaded regions: MGs, labeled by order $n$.
Markers at $k=4$ and $24$~rad/$\mu$m: cases studied in Fig.~\ref{fig:validation}. Only $\operatorname{Re}f_F\geq0$ is shown.
Parameters: $M_s=1.2$~MA/m, $A_{\mathrm{ex}}=15$~pJ/m, $d=1$~nm.
}
\label{fig:concept}
\vspace{-10pt}
\end{figure}

\paragraph*{Model.}
We consider an ultrathin ferromagnetic film of thickness $d$, with saturation magnetization $M_s$, exchange stiffness $A_{\mathrm{ex}}$, and PMA, magnetized in-plane along $y$ by a bias field $B_0$. The spin waves propagate along $x$, i.e., in the Damon--Eshbach geometry [Fig.~\ref{fig:concept}(a)].
Since the temporal impedance is set by the precession ellipticity~\cite{SobuckiGruszecki2026}, we require a system in which this ellipticity can be strongly tuned.
The competition between PMA and shape anisotropy provides this tunability near the spin-reorientation transition~\cite{Kisielewski2023, Lesniewski2026}.
On the stable in-plane side of this transition, small changes in PMA or applied field produce large changes in precession ellipticity~\cite{Lesniewski2026}.
This yields a large impedance contrast for a small change of the control parameter, while the ultrathin geometry allows voltage-controlled magnetic anisotropy (VCMA)~\cite{Chen2017,Deka2022}.
Spatially uniform periodic modulation of the reduced PMA, $Q(t)=2K_u(t)/(\mu_0M_s^2)=Q(t+T)$, conserves $k$ while allowing energy exchange with the drive.
Here, $K_u$ is the uniaxial anisotropy constant and $\mu_0$ the vacuum permeability.
For the protocol in Fig.~\ref{fig:concept}(a), the anisotropy switches between $Q_1=0.75$ and $Q_2=0.85$, i.e., with modulation depth $\Delta Q=Q_2-Q_1$, with period $T=0.1$~ns and equal durations of both states.
For these parameters, $\Delta Q=0.1$ corresponds to an interfacial anisotropy change of approximately $0.09$~mJ/m$^2$ (requiring electric fields of $0.26$--$0.65$~V/nm using VCMA coefficients of $350$--$140$~fJ/(V\,m) reported for VCMA switching~\cite{Nozaki2020,Nozaki2023_CoFeB}).

In the linear regime, the transverse components of dynamic magnetization $\mathbf m_k=(m_x,m_z)^{\mathsf T}$ obey the linearized  Landau–Lifshitz equation
\begin{equation}
    \partial_t\mathbf m_k=\Lmat\,\mathbf m_k,\quad
    \Lmat=\begin{pmatrix}
        0 & \omega_x(k,t)\\
        -\omega_z(k) & 0
    \end{pmatrix}
    +\ii\omega_D(k)\Id.
    \label{eq:linear_model}
\end{equation}
The coefficients $\omega_x$ and $\omega_z$ include the external field and the exchange, dipolar, and anisotropy contributions; their explicit forms are given in the End Matter.
The modulation of $Q$ changes only $\omega_x$, while $\omega_z$ remains constant.
We assume $\omega_x(k,t)>0$ and $\omega_z(k)>0$ throughout the cycle, which ensures a real precession frequency without DMI and the stability of the uniform state.
DMI enters through the term $\omega_D(k)=2\gamma Dk/M_s$, odd in $k$, where $D$ is the interfacial DMI constant, $\gamma>0$ is the modulus of the gyromagnetic ratio, and the sign convention of $D$ is fixed by Eq.~\eqref{eq:linear_model}.
Below we derive the lossless solution; Gilbert damping, with damping constant $\alpha$, is included in the End Matter, and the effect of DMI will be discussed later in this letter.

For constant $Q_i$, the precession frequency without DMI is $\omega_i=\sqrt{\omega_x^{(i)}\omega_z}$, and the ellipticity $Z_i=|m_z|/|m_x|=\sqrt{\omega_z/\omega_x^{(i)}}$ defines the temporal impedance~\cite{SobuckiGruszecki2026}.
The dispersion relations $f_i(k)=\omega_i(k)/(2\pi)$ of the two anisotropy states are shown as dashed lines in Fig.~\ref{fig:concept}(b,c).
Since the Landau--Lifshitz equation is first order in time, both magnetization components remain continuous at an anisotropy step.
For $Z_1\neq Z_2$, the incident mode splits into two modes at the same $k$ with opposite-sign frequencies, corresponding to time refraction and time reflection.
The time-reflected mode has reversed phase velocity and amplitude $r_{21}=(1-Z_1/Z_2)/2$ in $m_x$.
A periodic sequence of such interfaces thus repeatedly converts between the two modes, which underlies the band structure derived below.

\begin{figure}[!tb]
    \centering
    \includegraphics[width=\columnwidth,height=0.46\textheight,keepaspectratio]{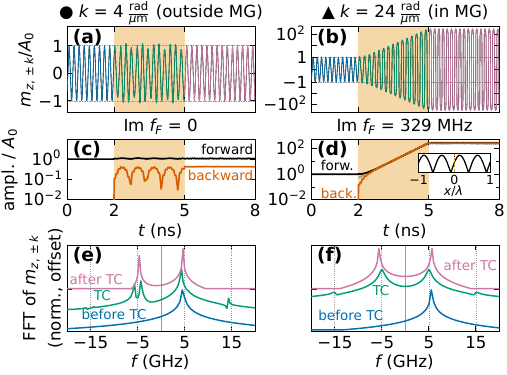}
    \vspace{-20pt}
    \caption{
Spin-wave dynamics outside ($k=4$~rad/$\mu$m, left) and inside ($k=24$~rad/$\mu$m, right) the first MG for $B_0=60$~mT; the modulation acts for $2\le t\le5$~ns (shading), and colors in (a,b,e,f) mark the intervals before, during, and after it.
(a,b) Contribution of the $\pm k$ pair to $m_z$ at $x=0$, $m_{z,\pm k}=2\operatorname{Re}a_k(t)$, where $a_k$ is the spatial Fourier amplitude of $m_z$ at $k$, normalized to its value $A_0$ before switch-on.
(c,d) Amplitudes of the forward and backward waves (positive- and negative-frequency components); dashed in (d): Floquet growth rate; inset: standing-wave profile $|m_z(x)|$ at the end of the modulation.
(e,f) Normalized spectra $|\tilde a_k(f)|$ in the three intervals (offset vertically); dotted: $nf_{\mathrm{mod}}/2$ with $n=1,2,\dots$; weaker peaks shifted by $\pm f_{\mathrm{mod}}$ from the main lines: Floquet sidebands.
} \label{fig:validation}
\vspace{-10pt}
\end{figure}

For a two-state modulation with durations $T_1$ and $T_2=T-T_1$, the monodromy matrix $\Mmat$ evolves the initial magnetization perturbation $\mathbf m_k(0)$ as $\mathbf m_k(nT)=\Mmat^n\mathbf m_k(0)$ after $n$ periods.
Its eigenvalues $\mu_\pm=\exp(-\ii\Omega_{F,\pm}T)$ define the Floquet quasifrequencies.
Composing the propagation in both states, with continuity at the interfaces, yields
\begin{equation}
    \begin{aligned}
        \cos\!\left[(\Omega_F+\omega_D)T\right]
        &=\Delta(k),\\
        \Delta(k)&=\cos\phi_1\cos\phi_2-\eta\sin\phi_1\sin\phi_2,
    \end{aligned}
    \label{eq:floquet_dispersion}
\end{equation}
where $\phi_i=\omega_iT_i$ and $\eta=(Z_1/Z_2+Z_2/Z_1)/2\geq1$, with equality only for $Z_1=Z_2$, in which case $\Delta=\cos(\phi_1+\phi_2)$ and no gaps open.
For $|\Delta|\leq1$ the quasifrequencies are real.
Wave-vector intervals with $|\Delta|>1$ form the MGs, in which one Floquet solution grows and the other decays at the same rate and, without DMI, $\operatorname{Re}f_F$ is pinned to integer multiples of $f_{\mathrm{mod}}/2$; each solution contains forward and backward waves, which therefore grow together [Fig.~\ref{fig:validation}(d)].
For the growing branch,
\begin{equation}
    \operatorname{Im}f_F(k)
    =
    \frac{1}{2\pi T}
    \operatorname{arccosh}|\Delta(k)|,
    \qquad
    f_F=\frac{\Omega_F}{2\pi},
    \label{eq:floquet_gain}
\end{equation}
and the amplitude grows as $\exp(2\pi\operatorname{Im}f_F\,t)$.

Figure~\ref{fig:concept}(b,c) show the resulting band structure for $B_0=60$ and $85$~mT, respectively; the shaded MGs correspond to nonzero $\operatorname{Im}f_F$ in the lower panels.
Increasing the field shifts the dispersions of both anisotropy states and thus the gap positions: the first-order gap, located at finite $|k|$ for 60~mT, extends to $k=0$ for 85~mT, where the mean frequency of the two states at $k=0$ reaches $f_{\mathrm{mod}}/2$.
Since the $k=0$ spin-wave frequency is set by the bias field, the field tunes the gap position continuously and in situ at a fixed modulation period, down to $k=0$, whereas in photonic time crystals this requires a cutoff fixed by design~\cite{Wang2025_bandgaps,MultiMode2026}.
The green symbols, obtained from separate micromagnetic simulations for each $k$, reproduce both the band structure and the growth rates in the MGs, including the narrow higher-order gaps.

Figure~\ref{fig:validation} follows the dynamics at the two wave vectors marked in Fig.~\ref{fig:concept}(b), outside ($k=4$~rad/$\mu$m) and inside ($k=24$~rad/$\mu$m) the first MG.
Outside the gap, the modulation generates a time-reflected component, visible in the spectra as a negative-frequency peak (reversed phase velocity), but the amplitudes remain bounded [Fig.~\ref{fig:validation}(a,c,e)].
Inside the gap, the forward and backward components grow exponentially to comparable amplitudes and form a standing wave [Fig.~\ref{fig:validation}(b,d), inset], analogous to photonic time crystals~\cite{Xiong2025}.
The growth rate agrees with $\operatorname{Im}f_F\simeq329$~MHz from the band structure and with the simulated value in Fig.~\ref{fig:concept}(b), and the spectral peaks during modulation lie at $\pm f_{\mathrm{mod}}/2$, where $\operatorname{Re}f_F$ is pinned at the Floquet zone edge [Fig.~\ref{fig:validation}(f)].
After the modulation is switched off, the amplified counterpropagating waves persist.

\begin{figure}[!tb]
  \centering
  \includegraphics[width=\columnwidth,height=0.46\textheight,keepaspectratio]{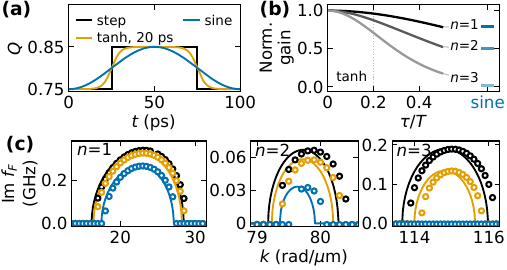}
  \vspace{-20pt}
  \caption{
        (a) Step, tanh ($\tau=20$~ps), and sinusoidal modulation for $B_0=60$~mT, $Q_1=0.75$, $Q_2=0.85$, and $T=100$~ps.
    (b) Maximum growth rate in each gap, normalized to the step: tanh versus $\tau/T$ (gray-scale curves, $n=1,2,3$; dotted line: $\tau=20$~ps) and sine (blue ticks at right); thin gray connectors link the results to the gap labels.
    (c) $\operatorname{Im}f_F(k)$ for the waveforms in (a): Floquet theory (lines) and micromagnetic simulations (open symbols); colors as in (a).
    }
\label{fig:sharpness}
\vspace{-10pt}
\end{figure}

\paragraph*{Modulation waveform.}
Since anisotropy switching in experiments has a finite rise time, we compare step, tanh, and sinusoidal modulation, the sine being the smoothest case [Fig.~\ref{fig:sharpness}; waveforms in the End Matter].
Smoothing reduces the growth rates and MG widths, more strongly at higher orders, yet tanh edges with $\tau=20$~ps ($\tau/T=0.2$) preserve a pronounced third gap, nearly suppressed by the sine, as confirmed micromagnetically [Fig.~\ref{fig:sharpness}(c)].
For weak modulation, the gap of order $n$ couples a wave at $f_n=nf_{\mathrm{mod}}/2$ to its time-reflected partner at $-f_n$ through the harmonic $q_n$ of $Q(t)$ at $nf_{\mathrm{mod}}$, with growth rate and width $\propto|q_n|$ (Supplemental Material).
For the 50\% step, $q_n=0$ for even $n$, whereas the odd gaps grow at comparable rates, since each sharp edge time-reflects the same fraction of the wave regardless of its frequency.
Tanh edges reduce $q_n$ and hence the maximum growth rate in each odd gap,
\begin{equation}
    \frac{\max_k\operatorname{Im}f_{F,n}^{\tanh}}{\max_k\operatorname{Im}f_{F,n}^{\mathrm{step}}}\simeq F_n=\frac{x_n}{\sinh x_n},\qquad x_n=\frac{\pi^2\tau f_n}{2},
    \label{eq:smoothing_ratio}
\end{equation}
where each maximum is taken within gap $n$; $F_n$ is set by the switching time relative to the wave period.
Since the sine contains only the fundamental, its gaps with $n\ge2$, like the even gaps, do not open at first order in $\Delta Q$, and $F_n$ does not apply to them.
Smoothed switching can thus preserve higher-order MGs with growth rates that exceed damping losses, enabling multiband amplification by a slow drive.

\begin{figure}[!tb]
  \centering
  \includegraphics[width=\columnwidth,height=0.46\textheight,keepaspectratio]{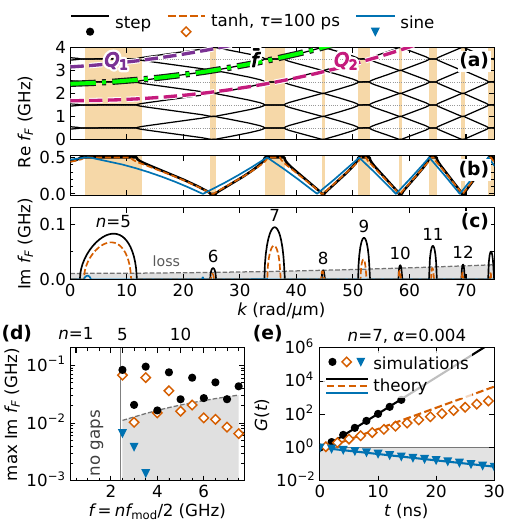}
  \vspace{-20pt}
  \caption{
Modulation at $f_{\mathrm{mod}}=1$~GHz ($Q_1=0.9$, $Q_2=1.0$, $B_0=60$~mT, 50\% duty cycle) for a step, tanh edges with $\tau=100$~ps, and a sine; other parameters as in Fig.~\ref{fig:concept}.
(a) $\mathrm{Re}\,f_F(k)$ of the step in the extended zone; dashed: unmodulated dispersions for $Q_1$ and $Q_2$; dash-dotted: their mean $\bar f$.
(b) $\mathrm{Re}\,f_F(k)$ in the reduced zone.
(c) $\mathrm{Im}\,f_F(k)$; numbers: gap order $n$.
(d) Maximum $\mathrm{Im}\,f_F$ of each gap versus $f_n=nf_{\mathrm{mod}}/2$; hatched: $f_n<\bar f(k=0)$.
(e) Gain $G(t)=|a_k(t)|/|a_k(0)|$, with $t=0$ at switch-on, at the center of the $n=7$ gap ($f_7=3.5$~GHz): monodromy (lines) and micromagnetics (symbols, shown up to $\max_x|m_z|=0.05$).
Shaded regions in (a)--(c): MGs of the step.
Dashed lines and gray areas in (c,d): Gilbert loss rate $\alpha\langle\omega_x+\omega_z\rangle/4\pi$ for $\alpha=0.004$, also used in (e).
  }
\label{fig:slow}
\vspace{-15pt}
\end{figure}

\paragraph*{Slow drive and a comb of high-order gaps.}
So far the studied waves above lay in the first gap, $f\approx f_{\mathrm{mod}}/2$.
Since the gap of order $n$ lies where the mean dispersion $\bar f=(f_1+f_2)/2$
reaches $f_n=nf_{\mathrm{mod}}/2$ [Fig.~\ref{fig:slow}(a)], a drive with
$f_{\mathrm{mod}}<2\bar f(0)$ cannot open the first gap and amplifies only
through gaps of order $n>2\bar f(0)/f_{\mathrm{mod}}$; we call such a drive slow.
For $f_{\mathrm{mod}}=1$~GHz, $Q_1=0.9$, and $Q_2=1$, $\bar f(0)\approx2.42$~GHz,
so the lowest gap has $n=5$.
For the stepwise drive, the odd gaps remain strong up to high orders, whereas the even ones are much weaker [Fig.~\ref{fig:slow}(c)], so that the odd gaps form a comb of gain bands spaced by $f_{\mathrm{mod}}$.
This follows from the phase condition, Eq.~\eqref{eq:phase_condition} (End Matter).
Within one period, the wave can return to its branch either by two time refractions, accumulating the phase $\phi_1+\phi_2$, or by two time reflections, accumulating $\delta=\phi_1-\phi_2=\pi(f_1-f_2)/f_{\mathrm{mod}}$.
Their interference sets the gap strength, $\propto\cos^2(\delta/2)$ for odd and $\sin^2(\delta/2)$ for even $n$.
Here $f_1-f_2\simeq2f_{\mathrm{mod}}$, i.e., $\delta\simeq2\pi$, so the odd gaps behave as in the weak-modulation limit and the even ones nearly vanish; since $f_1-f_2$ varies only slowly with $k$, this holds across the whole comb (End Matter).
The $n=7$ gap grows about seven times faster than the Gilbert losses for $\alpha=0.004$, and net gain occurs in all odd orders from $n=5$ to $13$, i.e., between $2.5$ and $6.5$~GHz [Fig.~\ref{fig:slow}(d)].

With tanh edges of $\tau=100$~ps, the 1~GHz drive still amplifies above losses in three bands, at $2.5$, $3.5$, and $4.5$~GHz, whereas a sine of the same frequency does not overcome the losses in any gap [Fig.~\ref{fig:slow}(d)].
Edges need only to be short compared with the wave period, not the drive period (here a tenth of the latter) so a single slow drive amplifies several bands at once, selected through $f_{\mathrm{mod}}$ and $\Delta Q$ via the phase condition.

Figure~\ref{fig:slow}(e) shows the gain $G(t)$ at the center of the $n=7$ gap, including damping.
At switch-on, the peak wave amplitude is $|m_z|\approx4\times10^{-5}$, with $m_z$ normalized to $M_s$.
We follow the simulations until it reaches $0.05$, below which the growth remains linear and the nonlinear frequency shift does not detune the wave from the gap; this limits the observable gain to $G\sim10^3$.
The simulations reproduce the predicted growth for the step ($G\approx750$ after 14~ns, when this limit is reached) and the decay for the sine ($G\approx0.07$ after 30~ns); for tanh edges, the wave is also amplified, $G\approx8\times10^2$ after 30~ns.
The slow drive thus amplifies spin waves above the losses also with smoothed edges, here with $\tau f_7=0.35$.

\begin{figure}[!tb]
  \centering
  \includegraphics[width=\columnwidth,height=0.46\textheight,keepaspectratio]{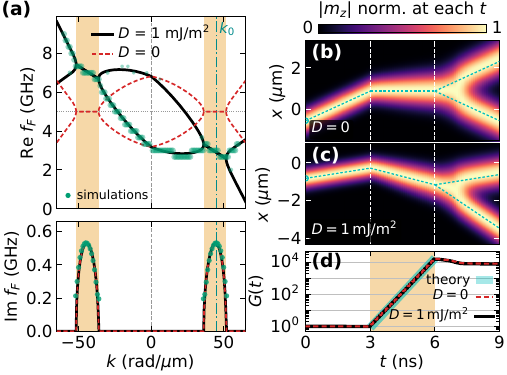}
  \vspace{-20pt}
  \caption{
(a) $\operatorname{Re}f_F$ (top) and $\operatorname{Im}f_F$ (bottom) for $D=1$~mJ/m$^2$ (black; dots: micromagnetic simulations) and $D=0$ (red dashed); shaded regions: MGs.
(b),(c) Envelope of $m_z(t,x)$ for a wave packet with $k_0=45$~rad/$\mu$m, for $D=0$ (b) and $D=1$~mJ/m$^2$ (c), normalized at each $t$ to its maximum over $x$; vertical dashed lines: modulation on and off; cyan: trajectories from the band structure at $k_0$.
(d) Envelope gain $G_{\mathrm{env}}(t)=m_z^{\max}(t)/m_z^{\max}(0)$, with $m_z^{\max}$ the maximum of the envelope over $x$, for both $D$; cyan band: slope $2\pi\operatorname{Im}f_F(k_0)$.
Parameters: $D=1$~mJ/m$^2$, $Q_1=0.85$, $Q_2=0.95$, $B_0=60$~mT, $T=0.1$~ns; others as in Fig.~\ref{fig:concept}.
} \label{fig:dmi}
  \vspace{-10pt}
\end{figure}


\paragraph*{In-gap drift of amplified wave packets.}
Interfacial DMI adds to $\Lmat$ in Eq.~\eqref{eq:linear_model} a term proportional to the identity, which enters Eq.~\eqref{eq:floquet_dispersion} only through $\Omega_F+\omega_D$ and thus shifts the quasifrequency by $-\omega_D(k)$ (End Matter).
In the lossless model, DMI therefore leaves the positions and widths of the MGs and the growth rates unchanged but tilts the real part of the quasifrequency [Fig.~\ref{fig:dmi}(a)].
Without DMI, $\operatorname{Re}\Omega_F$ is constant inside an MG and the group velocity vanishes; with DMI, gap modes acquire
\begin{equation}
    v_g=\frac{\partial\,\operatorname{Re}\Omega_F}{\partial k}
    =-\frac{\partial\omega_D}{\partial k}
    \equiv v_D=-\frac{2\gamma D}{M_s},
    \label{eq:dmi_group_velocity}
\end{equation}
which gives $v_D\approx-0.29$~$\mu$m/ns for $D=1$~mJ/m$^2$ and $M_s=1.2$~MA/m.
Since $\omega_D$ is linear in $k$, in the lossless linear model this holds for the entire field, $\mathbf m(x,t)=\mathbf m^{(0)}(x-v_Dt,t)$: the dynamics without DMI is translated with velocity $v_D$, so the $\pm k$ components that form the standing wave in the gap drift together.
DMI thus makes the film act as a time-modulated medium moving at $v_D$, without any spatial modulation.
Since the sign of $v_D$ follows that of $D$ relative to $\mathbf M_0$, reversing $B_0$ reverses the drift, so amplified packets can be steered out of the modulated region in either direction.

\paragraph*{Conclusions.}
We have established the magnonic counterpart of a photonic time crystal, a film with periodically switched PMA, deriving its Floquet band structure and confirming it micromagnetically.
MGs require an ellipticity contrast, which unifies temporal interfaces and parametric pumping, while the waveform harmonics and a phase condition set which high-order gaps open, providing design rules for the amplified bands.
As a result, a single 1~GHz drive amplifies a comb of spin-wave bands above Gilbert losses, with simulated gains approaching $10^3$, limited by nonlinearity.
Unlike spin--orbit-torque, stimulated, and parametric amplifiers~\cite{Merbouche2024,Breitbach2023,Nikolaev2025}, this drive lies below the signal frequencies and serves several bands at once.
DMI adds control of transport: amplified packets drift inside the gap without a change of the growth rate in the lossless model, and an adiabatic spin-transfer torque, which shifts the spin-wave frequency linearly in $k$~\cite{Vlaminck2008}, could tune this drift electrically.
The monodromy description applies to other periodic magnetic drives; trains of ultrashort spin--orbit-torque pulses~\cite{Janda2022UltrashortSOT} are a natural candidate, since their broad harmonic spectrum favors high-order gaps, their field-like component can modulate the ellipticity in a suitable geometry, and their damping-like component the balance of loss and gain.
Like any parametric amplifier, the drive also amplifies thermal magnons within the gaps, which can grow even without an input signal; the signal-to-noise ratio and the usable gain duration therefore have to be determined at finite temperature, which we leave for future work.
More broadly, magnonics offers momentum-gap physics beyond that of photonic time crystals: gaps tunable by a static field down to $k=0$ and, with DMI, amplified gap modes that propagate without spatiotemporal modulation, at GHz drive rates and anisotropy changes comparable to those in VCMA switching.

\begin{acknowledgments}
We acknowledge the use of the infrastructure of the Poznań Supercomputing and Networking Center (PCSS) within Scientific Computing Space no. pl0804.
\end{acknowledgments}

\section*{Data availability}
The data that support the findings of this article will be made openly available on Zenodo upon acceptance (DOI: \textcolor{red}{10.5281/zenodo.XXXXXXX}).

\bibliography{literature}

\clearpage

\section*{End Matter}
\label{sec:endmatter}
\setlength{\abovedisplayskip}{9pt plus 2pt minus 5pt}
\setlength{\belowdisplayskip}{8pt plus 2pt minus 5pt}
\setlength{\abovedisplayshortskip}{0pt plus 2pt}
\setlength{\belowdisplayshortskip}{5pt plus 2pt minus 2pt}

\noindent\textbf{\textit{Linear spin-wave model.}}---
Within the uniform-mode, thickness-averaged dipolar approximation appropriate for ultrathin films, the linearized Landau--Lifshitz equation for small deviations $m_{x,z}$ from $\mathbf M_0\parallel\hat y$ (neglecting damping) gives Eq.~\eqref{eq:linear_model} with~\cite{SobuckiGruszecki2026}
\begin{align}
 \omega_x&=A\,[h+\ell_{\rm ex}^2k^2+1-\xi(|k|d)-Q(t)],\notag\\
 \omega_z&=A\,[h+\ell_{\rm ex}^2k^2+\xi(|k|d)],
 \label{eq:em_components}
\end{align}
where $A=\gamma\mu_0M_s$, $h=B_0/(\mu_0M_s)$, $\ell_{\rm ex}^2=2A_{\rm ex}/(\mu_0M_s^2)$, and $\xi(u)=1-(1-e^{-u})/u$.
The reduced anisotropy is $Q(t)=2K_u(t)/(\mu_0M_s^2)$, where $K_u(t)=K_s(t)/d$ is the uniaxial anisotropy constant and $K_s$ the interfacial anisotropy per unit area.
We use $\gamma=176$~rad$\cdot$GHz/T, $M_s=1.2$~MA/m, $A_{\rm ex}=15$~pJ/m, and $d=1$~nm.
With the convention $e^{\ii kx-\ii\Omega t}$, interfacial DMI enters Eq.~\eqref{eq:linear_model} through $\omega_D=A\ell_Dk$ with $\ell_D=2D/(\mu_0M_s^2)$.

\noindent\textbf{\textit{Stepwise modulation and Floquet solution.}}---
The two-state protocol holds $Q_1$ and $Q_2$ for $T_1$ and $T_2$, with period $T=T_1+T_2=1/f_{\rm mod}$.
For constant $Q_i$, without DMI and damping, the precession frequency is $\omega_i=2\pi f_i=\sqrt{\omega_x^{(i)}\omega_z}$, the temporal impedance is $Z_i=|m_z|/|m_x|=\sqrt{\omega_z/\omega_x^{(i)}}$, and the propagator is
\begin{equation}
 \Umat_i(t)=\begin{pmatrix}
 \cos\omega_it&Z_i^{-1}\sin\omega_it\\
 -Z_i\sin\omega_it&\cos\omega_it
 \end{pmatrix}.
 \label{eq:propagator}
\end{equation}
Continuity of $(m_x,m_z)$ at the steps allows the propagators to be multiplied directly, giving the monodromy matrix without DMI, $\Mmat^{(0)}=\Umat_2(T_2)\,\Umat_1(T_1)$, with $\mathbf m_k(T)=\Mmat^{(0)}\mathbf m_k(0)$; the superscript $(0)$ denotes $D=0$ throughout.
Since $\det\Mmat^{(0)}=1$, its eigenvalues $\mu_\pm=\exp(-\ii\Omega_{F,\pm}^{(0)}T)$ satisfy $\mu^2-2\Delta\mu+1=0$ with $\Delta=\operatorname{Tr}\Mmat^{(0)}/2$, and the trace gives
\[
 \cos\bigl(\Omega_F^{(0)}T\bigr)=\Delta=\cos\phi_1\cos\phi_2-\eta\sin\phi_1\sin\phi_2,
\]
with $\phi_i=\omega_iT_i$ and $\eta=(Z_1/Z_2+Z_2/Z_1)/2$, i.e., Eq.~\eqref{eq:floquet_dispersion} without DMI.
DMI adds $\ii\omega_D\Id$ to $\Lmat$, so the propagator factorizes as $\Umat=e^{\ii\omega_Dt}\,\Umat^{(0)}$; this multiplies both eigenvalues by $e^{\ii\omega_DT}$ and shifts $\Omega_F=\Omega_F^{(0)}-\omega_D$, which gives the argument $(\Omega_F+\omega_D)T$ in Eq.~\eqref{eq:floquet_dispersion}; the factorization is exact for $\alpha=0$.
For $|\Delta|>1$, $|\mu_+|=|\Delta|+\sqrt{\Delta^2-1}=\exp(\operatorname{Im}\Omega_FT)$, which gives Eq.~\eqref{eq:floquet_gain} with $f_F=\Omega_F/2\pi$.
At fixed $k$, the Floquet solution has the form $\mathbf m_k(t)=e^{-\ii\Omega_Ft}\mathbf p_k(t)$ with $\mathbf p_k(t+T)=\mathbf p_k(t)$, so $\Omega_F$ is defined modulo $2\pi/T$ and the spectrum contains harmonics spaced by $f_{\rm mod}$.

\noindent\textbf{\textit{Arbitrary waveforms.}}---
We write $Q(t)=Q_1+\Delta Q\,w(t)$, with the period starting at the center of the $Q_1$ plateau and edge centers at $t_\uparrow=T_1/2$ and $t_\downarrow=T_1/2+T_2$.
For the step, $w=1$ between the edges and $0$ otherwise; for the sine, $w=[1-\cos(2\pi t/T)]/2$; for tanh edges, with $g(u)=\tfrac12\tanh(4u/\tau)$,
\begin{equation}
 w(t)=\sum_{p\in\mathbb Z}\bigl[g(t-t_\uparrow-pT)-g(t-t_\downarrow-pT)\bigr].
\end{equation}
An isolated edge completes $\tanh2\approx96\%$ of the change within $t_0\pm\tau/2$, and its 10--90\% time is $\tfrac12\operatorname{artanh}(0.8)\,\tau\simeq0.549\tau$.
Smooth profiles are sampled at the midpoints $t_j=(j-\tfrac12)\,\delta t$ of $N=256$ equal segments of duration $\delta t=T/N$, which define both the monodromy and the protocol used in the micromagnetic simulations,
\begin{equation}
 \Mmat=e^{\ii\omega_DT}\,\Mmat^{(0)},\quad
 \Mmat^{(0)}=\Umat_{Q(t_N)}(\delta t)\cdots\Umat_{Q(t_1)}(\delta t),
 \label{eq:profile_monodromy}
\end{equation}
where $\Umat_{Q(t)}$ is the propagator, Eq.~\eqref{eq:propagator}, at constant anisotropy $Q$; the product is exact for this piecewise-constant protocol.

Since each $\Umat_Q$ in Eq.~\eqref{eq:propagator} has unit determinant, $\det\Mmat^{(0)}=1$, and Eqs.~\eqref{eq:floquet_dispersion} and \eqref{eq:floquet_gain} hold for any waveform, with $\Delta=\operatorname{Tr}\Mmat^{(0)}/2$ evaluated numerically; equivalently, $\operatorname{Re}f_F=-\arg\mu/(2\pi T)$ and $\operatorname{Im}f_F=[\ln|\mu|]/(2\pi T)$, with $\mu$ the eigenvalues of $\Mmat$.
The eigenvalues of $\Mmat$ do not depend on the starting point of the period, and no weak-modulation approximation is involved.

\noindent\textbf{\textit{Phase condition for high-order gaps.}}---
For $T_1=T_2=T/2$, with $\Sigma=\phi_1+\phi_2$ and $\delta=\phi_1-\phi_2=\pi(f_1-f_2)/f_{\rm mod}$, Eq.~\eqref{eq:floquet_dispersion} becomes
\begin{equation}
 \Delta=\cos\Sigma-\frac{\eta-1}{2}\left(\cos\delta-\cos\Sigma\right),
 \qquad
 \eta-1=\frac{(Z_1-Z_2)^2}{2Z_1Z_2}.
\end{equation}
At the center of the gap of order $n$, $\Sigma=n\pi$, i.e., $(f_1+f_2)/2=nf_{\rm mod}/2$, and
\begin{equation}
\begin{gathered}
|\Delta_n|
= 1+\frac{\eta-1}{2}\left[1-(-1)^n\cos\delta\right]
\equiv 1+(\eta-1)\,P_n(\delta),\\
P_n(\delta)=
\begin{cases}
\cos^2(\delta/2), & n\ \text{odd},\\
\sin^2(\delta/2), & n\ \text{even}.
\end{cases}
\end{gathered}
\label{eq:phase_condition}
\end{equation}
The gap strength is thus the product of the impedance contrast $\eta-1$ and the phase factor $P_n(\delta)$.
For $f_1-f_2\ll f_{\rm mod}$ ($\delta\to0$), odd gaps are maximal and even gaps vanish; at strong modulation, odd gaps close for $f_1-f_2=(2m+1)f_{\rm mod}$ and are maximal for $f_1-f_2=2mf_{\rm mod}$, with the roles reversed for even gaps.
A gap closes when an integer number of precession half-periods fits into one plateau, $\sin\phi_i=0$, so that mode conversion at successive interfaces interferes destructively.
For $(\eta-1)P_n\ll1$, $\operatorname{arccosh}(1+\varepsilon)\simeq\sqrt{2\varepsilon}$ and Eq.~\eqref{eq:floquet_gain} gives the growth rate at the gap center,
\begin{equation}
 \operatorname{Im}f_F\simeq\frac{1}{2\pi T}\sqrt{2(\eta-1)\,P_n(\delta)}
 =\frac{f_{\rm mod}}{2\pi}\,\frac{|Z_1-Z_2|}{\sqrt{Z_1Z_2}}\,\sqrt{P_n(\delta)},
 \label{eq:gap_growth}
\end{equation}
with $Z_i$ and $\delta$ evaluated at the gap center.
The growth rate thus scales with $\sqrt{P_n}$, i.e., $|\cos(\delta/2)|$ or $|\sin(\delta/2)|$, so the odd--even contrast is weaker in the growth rate than in $P_n$.
For the parameters of Fig.~\ref{fig:slow}, Eq.~\eqref{eq:gap_growth} reproduces the maxima of $\operatorname{Im}f_F$ from the monodromy within 2\% for $n=5$--$13$.

The frequency difference follows exactly from $2\pi f_i = \omega_i=\sqrt{\omega_x^{(i)}\omega_z}$ as
\begin{equation}
 f_1-f_2=\frac{A\,\Delta Q}{2\pi}\,
 \frac{\sqrt{\omega_z}}{\sqrt{\omega_x^{(1)}}+\sqrt{\omega_x^{(2)}}},
 \label{eq:freq_difference}
\end{equation}
i.e., it is proportional to $\Delta Q$ and grows with the precession ellipticity $Z$.

For the parameters of Fig.~\ref{fig:slow}, $f_1-f_2$ grows slowly with $k$, from $1.47$~GHz at $k=0$ to about $1.6$ and $2.3$~GHz at the centers of the $n=5$ and $13$ gaps, so that $P_n$ stays above $0.6$ for the odd gaps $n=[5$--$13]$ and below $0.25$ for the even ones.
At the center of the $n=7$ gap ($k\simeq36$~rad/$\mu$m), $f_1-f_2\simeq2.03$~GHz, so that $\delta\simeq2\pi$ and $P_7\simeq1$; this gap reaches $\operatorname{Im}f_F\approx0.095$~GHz, compared with a Gilbert loss rate of $\approx0.014$~GHz for $\alpha=0.004$.
Reducing $\Delta Q$ to $\simeq0.06$ gives $f_1-f_2\simeq f_{\rm mod}$ and closes the $n=7$ gap, as confirmed by the monodromy also for tanh edges.
Under the conditions of Fig.~\ref{fig:sharpness}, $\delta/\pi\simeq0.1$--$0.2$ and $P_n\gtrsim0.9$ for odd $n$, so the weak-modulation factor $F_n$ (Supplementary Material) applies.

\noindent\textbf{\textit{Gilbert damping and net gain.}}---
Linearizing the Gilbert term around $\hat y$ adds $\alpha J\,\partial_t\mathbf m_k$ to the right-hand side of Eq.~\eqref{eq:linear_model}, with $J=\bigl(\begin{smallmatrix}0&1\\-1&0\end{smallmatrix}\bigr)$; since $J^2=-\Id$, $\partial_t\mathbf m_k=\Lmat_\alpha\mathbf m_k$, and the monodromy of the sampled protocol becomes
\begin{equation}
 \Lmat_\alpha=\frac{(\Id+\alpha J)\,\Lmat}{1+\alpha^2},\quad
 \Mmat=\tilde\Umat_{Q(t_N)}(\delta t)\cdots\tilde\Umat_{Q(t_1)}(\delta t),
 \label{eq:damped_generator}
\end{equation}
with $\tilde\Umat_Q(\delta t)=e^{\Lmat_\alpha(Q)\,\delta t}$, which for $\alpha=0$ equals $e^{\ii\omega_D\delta t}\,\Umat_Q(\delta t)$ and recovers Eq.~\eqref{eq:profile_monodromy}.
The net growth rate is $\operatorname{Im}f_F=\ln|\mu_{\max}|/(2\pi T)$, with $\mu_{\max}$ the eigenvalue of $\Mmat$ of largest modulus; the same $\Mmat$ propagates the mode amplitudes for the gain in Fig.~\ref{fig:slow}(e).
Since $\operatorname{Tr}\Lmat_\alpha=[2\ii\omega_D-\alpha(\omega_x+\omega_z)]/(1+\alpha^2)$, $|\mu_+\mu_-|=|\det\Mmat|=\exp[-\alpha\langle\omega_x+\omega_z\rangle T/(1+\alpha^2)]$ exactly, for any waveform and $D$: damping lowers the mean of the two Floquet growth rates by $\alpha\langle\omega_x+\omega_z\rangle/(4\pi)$ to first order in $\alpha$, which is the reference loss rate in Fig.~\ref{fig:slow} and the decay rate of each mode in the unmodulated film for $D=0$.
For $D\neq0$, the DMI term $\ii\omega_D(\Id+\alpha J)/(1+\alpha^2)$ does not commute with the rest of $\Lmat_\alpha$ unless $Z=1$, so the factor $e^{\ii\omega_DT}$ in Eq.~\eqref{eq:profile_monodromy} no longer separates exactly.

\noindent\textbf{\textit{Micromagnetic simulations.}}---
We solve the full Landau--Lifshitz--Gilbert equation in MuMax$^3$\cite{Vansteenkiste2014} for a 1-nm-thick film with the parameters given above, discretized along $x$ only (one cell across $y$ and $z$), with periodic boundary conditions along $x$ and $y$.
Two types of systems are simulated.
For the band-structure sweeps (Figs.~\ref{fig:concept}--\ref{fig:slow}), each value of $k$ is a separate plane-wave simulation; the length $L_x=2\pi/\Delta k$ is chosen so that the allowed wave vectors $k_n=n\Delta k$ include the target $k$, which therefore spans an integer number of wavelengths, and the cell size ($\approx1.25$~nm) is adjusted to $L_x$ rather than fixed independently (e.g., $\Delta k=2$~rad/$\mu$m gives $L_x=\pi$~$\mu$m).
For the wave packets in Fig.~\ref{fig:dmi}, a longer domain ($L_x=6\pi$~$\mu$m, cells $\approx2.5$~nm) hosts a packet with carrier $k_0=45$~rad/$\mu$m.
After relaxation at $Q_1$, the waves are excited by a weak out-of-plane field tuned to the positive-frequency branch, with a Gaussian temporal envelope and, for packets, also a Gaussian spatial envelope; the dynamic magnetic field is switched off before the PMA modulation starts.
The modulation is applied as a spatially uniform, time-dependent PMA following the same discretized profiles as the monodromy, so that theory and simulations treat identical protocols.
We use $\alpha=10^{-4}$, except $\alpha=0.004$ in Fig.~\ref{fig:slow}(e), and neglect thermal fluctuations.

The analysis extracts the complex amplitude of the prescribed spatial mode from the simulated magnetization:
\begin{equation}
 m_z(x_j,t)
 \;\xrightarrow{\mathcal F_x}\;
 \widetilde m_z(k_n,t)
 \;\xrightarrow{k_n=k}\;
 a_k(t),
 \label{eq:mode_selection}
\end{equation}
where $\mathcal F_x$ denotes the spatial Fourier transform evaluated at each recorded time and $a_k(t)=\widetilde m_z(k,t)$ retains both amplitude and phase.
Since $m_z$ is real, $a_{-k}=a_k^*$, and the contribution of the selected $\pm k$ pair at $x=0$ is $2\operatorname{Re}a_k(t)$, shown in Fig.~\ref{fig:validation}(a,b) normalized to the incident-wave amplitude $A_0$ before switch-on.
The temporal spectrum $\tilde a_k(f)$ [Fig.~\ref{fig:validation}(e,f)] gives $\operatorname{Re}f_F$, and its parts at $f>0$ and $f<0$, i.e., components with positive and negative phase velocity, give the forward and backward amplitudes in Fig.~\ref{fig:validation}(c,d).
Since the linearized dynamics at fixed $k$ is two-dimensional and $T$-periodic, stroboscopic samples during the modulation obey $a_k(t_{\rm on}+t_j+mT)=c_+\mu_+^m+c_-\mu_-^m$ for every phase $t_j$ within the period, independently of the details of the model.
A two-exponential (matrix-pencil) fit of these samples thus yields the Floquet multipliers of the simulated system, the counterparts of the eigenvalues of $\Mmat$, and $\operatorname{Im}f_F=\ln|\mu_{\max}|/(2\pi T)$ [symbols in Figs.~\ref{fig:concept}, \ref{fig:sharpness}(c), and \ref{fig:dmi}(a)], including damping (net loss below about 1~MHz for $\alpha=10^{-4}$).
Unlike an exponential fit of the envelope, this is not biased by the transient conversion between forward and backward waves near the gap edges.
In Fig.~\ref{fig:slow}(e), $G$ is $|a_k|$ relative to its value at switch-on; for the packets in Fig.~\ref{fig:dmi}, $G$ follows from $m_z$ filtered around $k_0$.
Details of the excitation, multiplier extraction and acceptance criteria, and filtering, with step-by-step examples for $k=4$, $12$, and $24$~rad/$\mu$m, are given in the Supplemental Material.

\end{document}